# Merging satellite and gauge-measured precipitation using LightGBM with an emphasis on extreme quantiles

Hristos Tyralis[1], Georgia Papacharalampous[2], Nikolaos Doulamis[3], Anastasios Doulamis[4]

[1] Department of Topography, School of Rural, Surveying and Geoinformatics Engineering, National Technical University of Athens, Iroon Polytechniou 5, 157 80 Zografou, Greece (montchrister@gmail.com, hristos@itia.ntua.gr, https://orcid.org/0000-0002-8932-4997)

[2] Department of Topography, School of Rural, Surveying and Geoinformatics Engineering, National Technical University of Athens, Iroon Polytechniou 5, 157 80 Zografou, Greece (papacharalampous.georgia@gmail.com, https://orcid.org/0000-0001-5446-954X)

[3] Department of Topography, School of Rural, Surveying and Geoinformatics Engineering, National Technical University of Athens, Iroon Polytechniou 5, 157 80 Zografou, Greece (ndoulam@cs.ntua.gr, https://orcid.org/0000-0002-4064-8990)

[4] Department of Topography, School of Rural, Surveying and Geoinformatics Engineering, National Technical University of Athens, Iroon Polytechniou 5, 157 80 Zografou, Greece (adoulam@cs.ntua.gr, https://orcid.org/0000-0002-0612-5889)

**Abstract**: Knowing the actual precipitation in space and time is critical in hydrological modelling applications, yet the spatial coverage with rain gauge stations is limited due to economic constraints. Gridded satellite precipitation datasets offer an alternative option for estimating the actual precipitation by covering uniformly large areas, albeit related estimates are not accurate. To improve precipitation estimates, machine learning is applied to merge rain gauge-based measurements and gridded satellite precipitation products. In this context, observed precipitation plays the role of the dependent variable, while satellite data play the role of predictor variables. Random forests is the dominant machine learning algorithm in relevant applications. In those spatial predictions settings, point predictions (mostly the mean or the median of the conditional distribution) of the dependent variable are issued. The aim of the manuscript is to solve the problem of probabilistic prediction of precipitation with an emphasis on extreme quantiles in spatial interpolation settings. Here we propose, issuing probabilistic spatial predictions of precipitation using Light Gradient Boosting Machine (LightGBM). LightGBM is a boosting algorithm, highlighted by prize-winning entries in prediction and forecasting competitions. To assess LightGBM, we contribute a large-scale application that includes merging daily precipitation measurements in contiguous US with PERSIANN and GPM-IMERG satellite precipitation data. We focus on extreme quantiles of the probability distribution of the dependent variable, where LightGBM outperforms quantile regression forests (QRF, a variant of random forests) in terms of quantile score at extreme quantiles. Our study offers understanding of probabilistic predictions in spatial settings using machine learning.

## 1. Introduction

Economic constraints limit the extent as well as the density of spatial coverage of areas with rain gauge stations. Therefore, gridded satellite datasets are used as a substitute of observed precipitation in hydrological applications. Nevertheless, gridded satellite datasets provide inaccurate estimates of actual precipitation, therefore post-processing is required by merging gridded datasets with rainfall gauge-based measurements; see the reviews by Abdollahipour et al. (2022) and Hu et al. (2019).

A common means to merge gridded satellite datasets and gauge-based measurements is to apply machine learning algorithms in regression settings. In this context, satellite precipitation data play the role of predictor variables, while observed precipitation plays the role of the dependent variable. The major assumption in such settings is that the actual precipitation is represented by gauge-based measurements, albeit some studies question the accuracy of those measurements (Sun et al. 2018). The state-of-the-art algorithm in these regression settings is Breiman's (2001) random forests (Baez-Villanueva et al. 2020, Chen et al. 2021, Fernandez-Palomino et al. 2022, He et al. 2016, Hengl et al. 2018, Lin et al. 2022, Militino et al. 2023, Nguyen et al. 2021, Tyralis et al. 2019b, Zhang et al. 2021).

Other machine learning algorithms also have been implemented (e.g. Meyer et al. 2016, Lei et al. 2022, Papacharalampous et al. 2023a, 2023b, Shen and Yong 2021), albeit less frequently. To better understand their performance, it is important to compare multiple algorithms using big datasets that cover large areas with dense networks of rain gauge stations (Lei et al. 2022, Papacharalampous et al. 2023a, 2023b).

A common characteristic of most studies merging satellite data and station observations, is that spatial point predictions are issued and assessed using the squared error scoring function, the absolute error scoring function or related skill scores (e.g. the Nash-Sutcliffe efficiency (NSE) and the Kling-Gupta Efficiency (KGE)). The squared error scoring function is consistent (for the definition of consistency the reader is referred to Section 3.4) for the mean functional of the probability distribution (Gneiting 2011), i.e. by training a machine algorithm with a squared error scoring function, one can issue predictions of the mean of the conditional probability of the response of the regression algorithm. Similar arguments apply for the case of the absolute error scoring function (which is consistent for the median functional) as well as for the associated skill scores (Gneiting 2011). However, predictions are more informative when they take the form of probability distributions (Gneiting and Raftery 2007), while the requirement for probabilistic predictions in hydrology has been commented by Papacharalampous et al. (2019) and Papacharalampous and Tyralis (2022a) and has been identified as an important problem in hydrology, in the context of uncertainty estimation in general (Blöschl et al. 2019).

Prediction of quantiles of the conditional probability distribution at a dense grid of quantile levels can provide an approximation of the full probability distribution (Tyralis and Papacharalampous 2021b, 2022a). Our focus is on extreme quantiles of the conditional probability distribution (see e.g. Curceac et al. 2020, Tyralis and Papacharalampous 2023), although predictions of functionals in the centre of the conditional probability distribution of the response are also assessed. Extreme quantiles are of interest, given their importance for conducting flood studies. Our application includes merging the PERSIANN and the GPM-IMERG daily precipitation datasets with gauge-based daily precipitation measurements that cover the contiguous US (CONUS).

The aim of the manuscript is to solve the problem of probabilistic prediction of precipitation with an emphasis on extreme quantiles in spatial interpolation settings. To this end, we propose to apply the Light Gradient Boosting Machine (LightGBM) algorithm (Ke 2017) trained with the quantile scoring function (Koenker and Bassett Jr 1978). Applications of the quantile scoring function and the associated mean score can be found in hydrological modelling and forecasting studies, see e.g. Dogulu et al. 2015, Papacharalampous et al. 2019, Tyralis and Papacharalampous 2021b. LightGBM is compared with quantile regression forests (QRF, Meinshausen 2006), which is a variant of random forests. The assessment of LightGBM is based on its relative performance with respect to quantile regression forests, given the strong preference for the use of random forests in spatial interpolation settings due to high predictive performance and convenience in their use (Hengl et al. 2018). Previous applications of QRF in merging satellite and gauge-based precipitation measurements include Bhuiyan et al. (2018), Zhang et al. (2022), while applications of other machine learning algorithms that can issue probabilistic predictions in spatial interpolation settings are limited (see e.g. a deep learning application by Glawion et al. 2023). As already mentioned, we focus on extreme quantiles while we base the comparison on skill scores whose components include quantile scoring functions. Although the assessment focuses on extreme quantiles, we further comment on quantiles at central quantile levels, where results are largely influenced by the intermittent nature of precipitation.

The remainder of the manuscript is structured as follows. Section 2 presents LightGBM and QRF with emphasis on topics related to quantile regression. Section 3 follows with presentation of the data used in the study as well as the problem formulation and metrics for the assessment of the algorithms. Results are presented in Section 4, followed by their discussion in Section 5. The manuscript concludes with Section 6.

## 2. Methods

We applied Light Gradient Boosting Machine (LightGBM) to a dataset that includes gridded satellite precipitation products (see Section 3.1.2) and gauge-based measurements (see Section 3.1.1). Since our scope is to provide probabilistic predictions of precipitation (in particular high quantiles), LightGBM was trained using the quantile scoring function (see Section 2.3). The algorithms were compared with QRF using a hold-out sample (see Section 3.2). In this Section we describe LightGBM and QRF, while software implementation is provided in Appendix A. Since LightGBM and QRF are variants of boosting algorithms and random forests respectively, we focus on certain properties that distinguish them from the respective introducing algorithms. Extended descriptions of boosting and random forests can be found in textbooks (Hastie et al. 2009; James et al. 2013; Efron and Hastie 2016) while remote sensing scientists and technologists are familiar with them.

## 2.1 Quantile regression forests

Quantile regression forests (Meinshausen 2006) is a variant of random forests (Breiman 2001) that is used to issue probabilistic predictions. Random forests is a state-of –the-art algorithm for spatial interpolation (Hengl et al. 2018) and has been extensively used in hydrological applications (Tyralis et al. 2019b).

A random forests algorithm for regression grows an ensemble of decision trees while the prediction of the algorithm is equal to the mean of the individual trees. Building the forests of trees is done with bagging combined with randomized node optimization. Bagging (bootstrap aggregating) refers to the procedure of resampling with replacement of the training set, and using this sample to train a single tree. In addition, random forests select a random subset of features at each candidate split.

Quantile regression forests define an approximation of the conditional distribution of the response variable instead of averaging predictions of trees (a procedure that approximates the conditional mean of the response variable). Properties of random forests are transferrable to quantile regression forests. Those are summarized in Tyralis et al. (2019b) and include among others related to our problem at hand, high predictive performance, speed, feasibility in large scale applications, resistance to overfitting, efficient handling of highly correlated variables and stability.

Here we applied the R language implementation of quantile regression forests by Wright 2022 and Wright and Ziegler 2017), using 100 decision trees and default values of hyperparameters, since default implementation is highly efficient (Tyralis et al. 2019b). The specific software implementation of random forests is fast regarding computations times; however it is extremely slower compared to LightGBM for big datasets, thus hyperparameter optimization becomes prohibitive for the large sample of the present study. Since quantile regression forests are a regression algorithm, they are trained and predict in the usual fashion, i.e. the training sample includes a set of observed predictor variables and a set of the observed dependent variable. The specific application is described later in Section 3.2.

## 2.2 Light gradient boosting machine (LightGBM)

Gradient boosting decision trees is an ensemble learning algorithm in which decision trees are added to the ensemble sequentially. At each iteration, a new decision tree is trained with respect to the error of the algorithm so far. A gradient-descent based formulation formalized the concept of boosting (Friedman 2001, Natekin and Knoll 2013, Mayr et al. 2014). Gradient boosting decision trees can be optimized with different scoring functions, thus they can issue predictions tailored to user's requirements. A list of properties of boosting algorithms can be found in Tyralis and Papacharalampous (2021a). Although boosting algorithms share some similar properties with random forests, they frequently perform better in several settings, although hyperparameter tuning is needed.

LightGBM (Ke et al. 2017) is a boosting algorithm that has some favourable properties compared to common gradient boosting algorithms. In particular, it is particularly suited for datasets with high feature dimension and large size. It uses gradient-based on-side sampling (GOSS) that excludes data instances with small gradients (instead common boosting algorithms scan all data instances to estimate the information gain of all possible split points), thus reducing training time. Furthermore, it uses Exclusive Feature Bundling (EFB) that bundles mutually exclusive features, to reduce their number. Besides, it uses a histogram-based algorithm to find the best split points, similarly to earlier successful boosting variants (e.g. XGBoost, Chen and Guestrin 2016).

LightGBM has multiple parameters that when tuned can increase its predictive performance. The optimization procedure is described in Section 3.3. Moreover, LightGBM was trained with a quantile scoring function (see Sections 2.3 and 3.4) to issue probabilistic predictions. Since LightGBM is a regression algorithm, it is trained and predicts in the usual fashion, i.e. the training sample includes a set of observed predictor variables and a set of the observed dependent variable. The specific application is described later in Section 3.2.

## 2.3 Quantile regression

Quantile regression algorithms are used to predict conditional quantiles in regression settings. Here we explain how quantile regression works in practice. Hereinafter, observations will be notated with lowercase letters, while random variables will be notated by underlined lowercase letters. Let $\underline{y}$ be a random variable with cumulative distribution function $F_{\underline{y}}$ defined by:

$$F_y(y) := P(y \leq y) \tag{1}$$

Then, the $\tau^{\text{th}}$ quantile of $y$, $Q_y(\tau)$ is defined by:

$$Q_y(\tau) := \inf\{y: F_y(y) \geq \tau\}, \tau \in (0, 1) \tag{2}$$

where inf{·} denotes the infimum of a set of real numbers.

Let $F_{y|x}$ be the distribution of the random variable $y$ given the $p$-dimensional vector $\boldsymbol{x}$:

$$F_{y|x}(y|\boldsymbol{x}) := P(y \leq y|\boldsymbol{x}) \tag{3}$$

Then, the $\tau^{\text{th}}$ quantile of $y$ conditional on $\boldsymbol{x}$, $Q_{y|x}(\tau|\boldsymbol{x})$ is defined by:

$$Q_{y|x}(\tau|\boldsymbol{x}) := \inf\{y: F_y(y|\boldsymbol{x}) \geq \tau\}, \tau \in (0, 1) \tag{4}$$

The quantile loss function $\rho_\tau(u)$ is defined by eq. (5):

$$\rho_\tau(u) := u\,(\mathbb{I}(u \geq 0) - \tau), u \in \mathbb{R} \tag{5}$$

Here $\tau$ is the quantile level of interest and $\mathbb{I}(A)$ denotes the indicator function that is equal to 1 when the event $A$ realizes and 0 otherwise. The quantile loss function, defined by eq. (5), is positive and negatively oriented, i.e. the objective is to minimize it, and equals to 0, when $u = 0$.

Let $\boldsymbol{\theta}$ be the parameters of a regression model (e.g. the LightGBM of Section 2.2). Let $y(\boldsymbol{x}, \boldsymbol{\theta}(\tau))$ be the prediction of the regression model at quantile level $\tau$, given values of the predictor variables equal to $\boldsymbol{x}$, and values of the model's parameters equal to $\boldsymbol{\theta}(\tau)$. To estimate $\boldsymbol{\theta}(\tau)$ for $\tau \in (0, 1)$, one should minimize the average quantile score $(1/n) \sum_{i=1}^{n} \rho_\tau(y(\boldsymbol{x}_i, \boldsymbol{\theta}(\tau)) - y_i)$, which is the core idea of linear-in-parameters quantile regression elaborated by Koenker and Bassett Jr (1978). The regression model with parameters $\boldsymbol{\theta}(\tau)$ predicts conditional quantiles $Q_{y|x}(\tau|\boldsymbol{x})$.

## 3. Data and application

### 3.1 Data

We assessed the algorithms using daily earth-observed precipitation retrieved from the Global Historical Climatology Network daily (GHCNd) as described in Section 3.1.1, gridded satellite precipitation from the current operational PERSIANN (Precipitation Estimation from Remotely Sensed Information using Artificial Neural Networks) system as well as the GPM IMERG (Integrated Multi-satellitE Retrievals) late Precipitation dataset (described in Section 3.1.2) and elevation data retrieved from the Amazon Web Services (AWS) Terrain Tiles application (described in Section 3.1.3). The locations of gauges are presented in Figure 1.

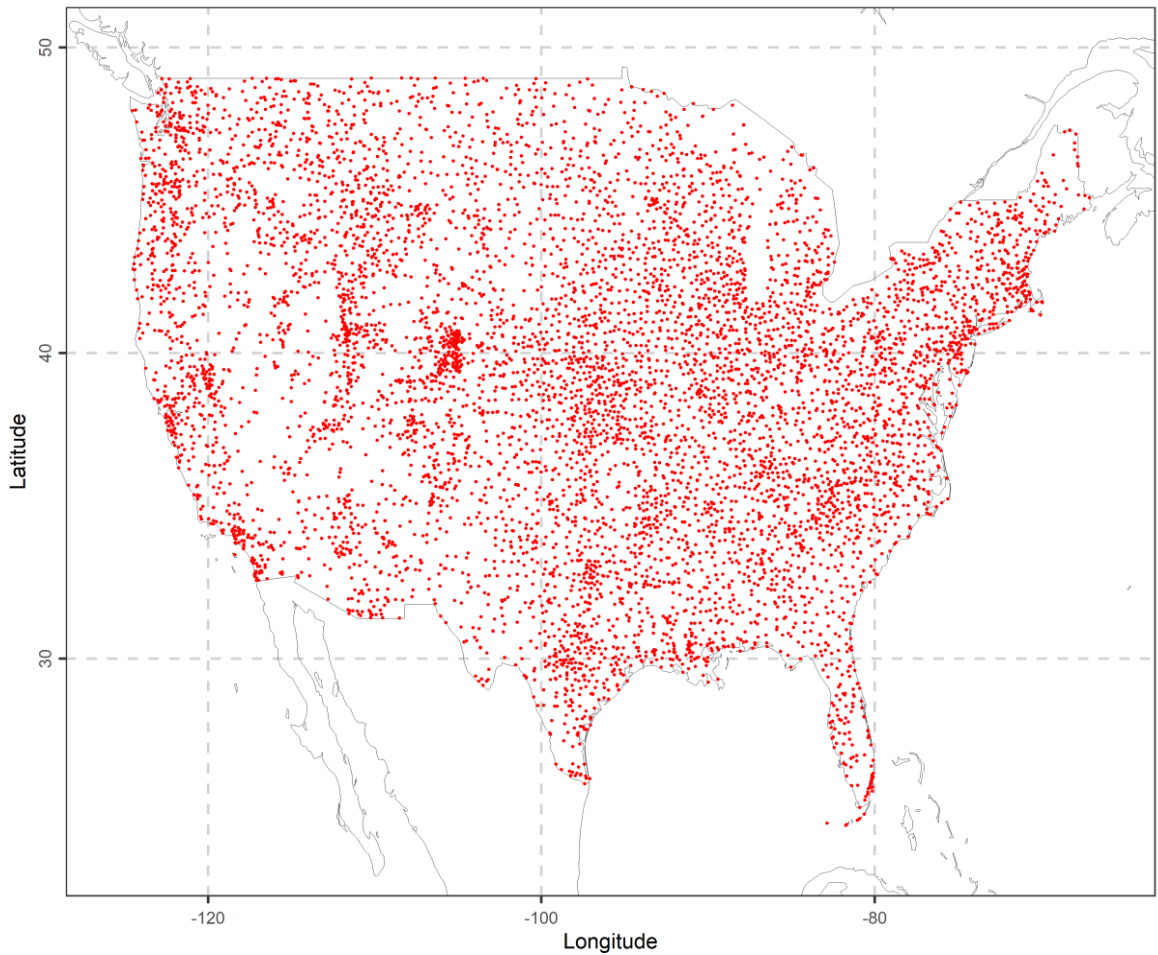

Figure 1. Map of the geographical locations (red points) of the earth-located stations that offered data for this work. Here, latitudes measure distance north of the equator and longitudes measure distance west of the meridian in Greenwich, England in degrees (°).

*3.1.1 Earth-observed precipitation data*

We used total daily precipitation data precipitation retrieved from the Global Historical Climatology Network daily (GHCNd) (Durre et al. 2008, 2010, Menne et al. 2012, NOAA National Climatic Data Center (https://www1.ncdc.noaa.gov/pub/data/ghcn/daily); accessed on 2022-02-27). In particular, data from 7 261 stations spanning across CONUS (see Figure 1) were extracted. The data cover the two-year time period 2014–2015.

*3.1.2 Satellite precipitation data*

We used gridded satellite daily precipitation data from the current operational PERSIANN system (Hsu et al. 1997, Nguyen et al. 2018, 2019), developed by the Centre for Hydrometeorology and Remote Sensing (CHRS) at the University of California, Irvine (UCI). The PERSIANN data were retrieved from the website of the Center for Hydrometeorology and Remote Sensing (CHRS) (https://chrsdata.eng.uci.edu; accessed on 2022-03-07).

Furthermore, we used the GPM IMERG late Precipitation L3 1 day 0.1 degree x 0.1 degree V06 dataset developed by the NASA Goddard Earth Sciences (GES) Data and Information Services Center (DISC) (Huffman et al. 2019). The GPM IMERG data were interpolated to a 0.25 degree x 0.25 degree using bilinear interpolation on CMORPH0.25 grid. Alternatives for interpolating exist, but conclusions of the manuscript are independent of the type of interpolation while they depend on the type of algorithm implemented (as explained later in Section 5). The IMERG data were retrieved from the website of NASA (National Aeronautics and Space Administration) Earth Data (https://doi.org/10.5067/GPM/IMERGDL/DAY/06; accessed on 2022-12-10).

The extracted data cover the contiguous United States at the two-year time period 2014–2015. The herein used GPM IMERG version of satellite precipitation product has not used ground-based precipitation data for bias correction. The PERSIANN dataset has been corrected using ground-based data; therefore applying a regression algorithm to the data is practically a post-processing framework that further improves the satellite dataset. That is a common approach in the field, e.g. see Baez-Villanueva et al. (2020) and Bhuiyan et al. (2018) among others.

It is possible to use different or more satellite precipitation datasets. In the latter case the available information and the number of predictor variables will increase followed by improved accuracy of the corrected precipitation product. Although, using multiple satellite precipitation datasets is a common practice when building new datasets which must be accurate, that is out of the scope of the present study, which aims to provide an understanding of the properties of LightGBM when probabilistic predictions are issued. Furthermore, conclusions (see Section 6) will not be affected, since results are due to theoretical properties of the two implemented algorithms as discussed later in Section 5. Nevertheless, including two datasets from which the one is already corrected, but the other is not, has an additional advantage. In particular, as already shown in Papacharalampous et al. (2023b), the uncorrected dataset includes more information compared to the corrected one (in fact the latter provides less significant improvements regarding the accuracy of the final product). Therefore, diversity of the products might serve to understand better the properties of the correction algorithms.

*3.1.3 Elevation data*

Elevation is a useful predictor variable when merging gauged-based and satellite precipitation data (Xiong et al. 2022). Therefore, we computed the elevation of the stations in Section 3.1.1 using the Amazon Web Services (AWS) Terrain Tiles (https://registry.opendata.aws/terrain-tiles; accessed on 2022-09-25) application.

## 3.2 Problem formulation and assessment of the algorithms

The setting of the problem has been formulated as follows similarly to the procedures proposed by Papacharalampous et al. (2023a, b). The total daily station precipitation is the dependent variable in a regression problem. Predictor variables are the total daily precipitations from the closest grid points to the station. In particular, there are four predictor variables corresponding to the PERSIANN dataset and another four predictor variables corresponding to the GMP IMERG dataset. Furthermore, we computed the distances between the station and the closest grid points for each satellite dataset; thus we obtained eight more predictor variables. The station elevations and their longitude and latitude also play the role of predictor variables. Possible interactions between predictor variables do not affect the performance of random forests (and their variants) as well as boosting (and its variants) as explained earlier in Sections 2.1 and 2.2 respectively.

To understand how we related station precipitation to the gridded satellite precipitation we designed Figure 2. For a single grid (e.g. the PERSIANN), we determined the closest four grid points to each precipitation station, we computed the distances $d_i$, $i = 1, 2, 3, 4$ from these grid points and ordered in increasing order $d_1 < d_2 < d_3 < d_4$ (see Figure 2). When we refer to the PERSIANN dataset, the distances $d_i$, $i = 1, 2, 3, 4$ are called "PERSIANN distances 1–4" while when we refer to the IMERG dataset, the distances are called "IMERG distances 1–4". The respective daily precipitation values at the grid points 1–4 are called "PERSIANN values 1–4" or "IMERG values 1–4".

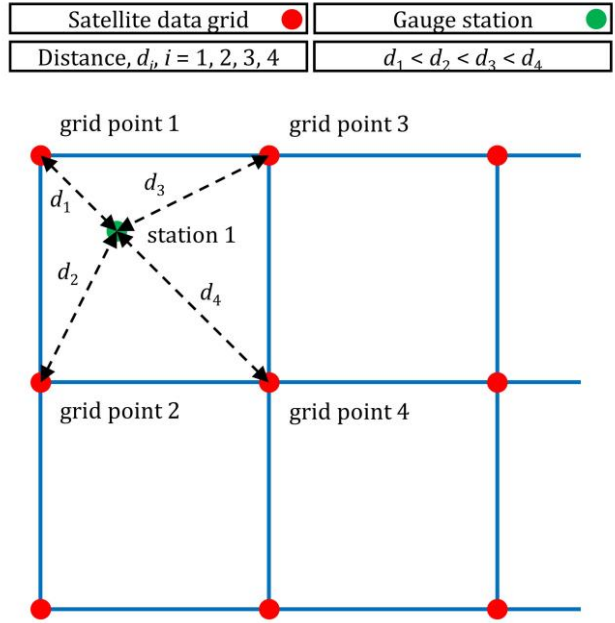

Figure 2. Setting of the regression problem. Note that the term "grid point" is used to describe the geographical locations with satellite data, while the term "station" is used to describe the geographical locations with ground-based measurements. Note also that, throughout this work, the distances $d_i$, $i$ = 1, 2, 3, 4 are also respectively called "PERSIANN distances 1–4" or "IMERG distances 1–4" (depending on whether we refer to the PERSIANN grid or the IMERG grid) and the daily precipitation values at the grid points 1–4 are called "PERSIANN values 1–4" or "IMERG values 1–4" (depending on whether we refer to the PERSIANN grid or the IMERG grid).

Table 1 presents the predictor variables for the regression setting of the problem. In particular, the predictor variables are the PERSIANN values 1–4, the IMERG values 1–4, the PERSIANN distances 1–4, the IMERG distances 1–4 and the station's longitude, latitude and elevation. The final dataset includes 4 833 007 samples. We split the dataset into three equally sized folds randomly. Random forests were trained in the union of the first two folds and were tested in the third fold (that includes 1 611 002 samples). LightGBM was trained in the first fold and was validated in the second fold. We implemented this procedure to estimate its hyperparameters (see Section 3.3). After estimating the LightGBM's hyperparamaters we re-trained the algorithm using the final parameters in the union of the first two folds. The algorithm was then tested in the third fold. Predictions of LightGBM lower than 0 were transformed to 0 (it is well known that precipitation is a positive quantity), while QRF did not issue negative precipitation predictions.

Table 1. Inclusion of predictor variables in the predictor sets examined in this work.

| Predictor variable | Predictor set |
|---|---|
| PERSIANN value 1 | ✓ |
| PERSIANN value 2 | ✓ |
| PERSIANN value 3 | ✓ |
| PERSIANN value 4 | ✓ |
| IMERG value 1 | ✓ |
| IMERG value 2 | ✓ |
| IMERG value 3 | ✓ |
| IMERG value 4 | ✓ |
| PERSIANN distance 1 | ✓ |
| PERSIANN distance 2 | ✓ |
| PERSIANN distance 3 | ✓ |
| PERSIANN distance 4 | ✓ |
| IMERG distance 1 | ✓ |
| IMERG distance 2 | ✓ |
| IMERG distance 3 | ✓ |
| IMERG distance 4 | ✓ |
| Longitude | ✓ |
| Latitude | ✓ |
| Station elevation | ✓ |

## 3.3 LightGBM hyperparameter optimization

LightGBM has multiple parameters that can be tuned. We selected to optimize some of them, while we kept the default values in the R software implementation (Shi et al. 2022) for the remaining parameters. The selection of hyperparameters to be tuned was directed by the algorithm's documentation as well as the experience in practical applications, since the algorithm has been part of prize-winning solutions in international prediction competitions. The parameters selected for tuning, along with their description based on software's documentation (https://lightgbm.readthedocs.io/en/v3.3.2/Parameters.html) follows in Table 2.

Table 2. LightGBM parameters.

| Parameter | Description | Values |
|---|---|---|
| `max_depth` | Max depth for tree model. `max_depth` can be used to limit the tree depth explicitly. | 6, 8, 10 |
| `min_data_in_leaf` | This is a very important parameter to prevent over-fitting in a leaf-wise tree. Its optimal value depends on the number of training samples and `num_leaves`. Setting it to a large value can avoid growing too deep a tree, but may cause under-fitting. In practice, setting it to hundreds or thousands is enough for a large dataset | 20, 100, 200, 500, 1 000 |
| `learning_rate` | Shrinkage rate. As a general rule, if one reduces `num_iterations`, then he should increase `learning_rate` | 0.02, 0.05, 0.1 |
| `num_iterations` | Number of iterations. The `num_iterations` parameter controls the number of boosting rounds that will be performed. Since LightGBM uses decision trees as the learners, this can also be thought of as "number of trees" | 400 |
| `num_leaves` | Max number of leaves in one tree. This is the main parameter to control the complexity of the tree model | 20, 40, 60, 80, 100, 200, 500 |

The parameter space includes a grid with all possible combinations of parameter's values, excluding set of parameters where `num_leaves` > $2^{\text{max\_depth}}$. Furthermore, we applied the algorithms with early stopping, setting the parameter `early_stopping_round` equal to 20. In this case the algorithm stops before reaching the specified number of iterations, if for 20 iterations there is no improvement in the score. Early stopping serves in reducing training time.

### 3.4 Performance metrics and assessment

We compared quantile regression forests and LightGBM using the quantile scoring function defined by:

$$S_\tau(x, y) := \rho_\tau(x - y) \qquad (6)$$

Here $y$ is the materialization (observation) of the spatial process and $x$ is the predictive quantile at level $\tau$. Hydrological predictions should be probabilistic in nature taking the form of probability distributions (see e.g. Gneiting and Raftery 2007, Papacharalampous and Tyralis 2022a). Predicting quantiles of the probability distribution at multiple levels is a nice substitute of the full probability distribution. The quantile scoring function is strictly consistent for the quantile functional of the predictive distribution, in the sense that if one receives a directive to predict a quantile, the expected quantile score is minimized when following the directive (Gneiting 2011). Therefore, when receiving a directive to predict a quantile functional, it is natural to train a model using the quantile scoring function as already mentioned in Section 2.3.

The performance criterion for the machine learning algorithms at quantile level $\tau$ takes the form:

$$\bar{S}_\tau := (1/n) \sum_{i=1}^{n} S_\tau(x_i, y_i) \qquad (7)$$

where $\{x_i, y_i\}$, $i = 1, ..., n$ are the predictions and observations for the $i^{\text{th}}$ sample and $n$ is the size of the test fold. We computed $\bar{S}_\tau$ at several quantile levels $\tau$ and for both algorithms. We considered predictions issued by quantile regression forests as reference predictions and we computed a skill score for the reference algorithm at the specified quantile level defined by:

$$S_{\tau,\text{skill}} := 1 - \bar{S}_{\tau,\text{LightGBM}} / \bar{S}_{\tau,\text{QRF}} \qquad (8)$$

In general, $S_{\tau,\text{skill}} \leq 1$, while for an excellent forecast at level $\tau$, we have $\bar{S}_{\tau,\text{LightGBM}} = 0$ and $S_{\tau,\text{skill}} = 1$. When $S_{\tau,\text{skill}} > 0$, LightGBM outperforms quantile regression forests, while the higher the $S_{\tau,\text{skill}}$, the better the LightGBM. We did not compare the algorithms using alternative scoring functions (e.g., the squared error scoring function, or a related skill score, e.g. the Nash-Suttcliffe efficiency) because such functions are not consistent for the quantile functional (Gneiting 2011).

In addition, we computed a score for each algorithm that characterizes how well each algorithm issues predictions with the nominal frequencies. In particular, the respective score is defined by

$$\overline{\text{FR}}_\tau := |(1/n) \sum_{i=1}^{n} \mathbb{I}(y_i \leq x_i) - \tau| \qquad (9)$$

To better understand the score, let $\tau = 0.95$. Assuming that perfect predictions have been issued, then $(1/n) \sum_{i=1}^{n} \mathbb{I}(y_i \leq x_i)$ should be equal to 0.95 (i.e. 95% of observations should be lower or equal to respective predictions) and $\overline{\text{FR}}_\tau$ should be equal to 0. Again, we computed the respective skill score, with quantile regression forests as reference algorithm:

$$\text{FR}_{\tau,\text{skill}} := 1 - \overline{\text{FR}}_{\tau,\text{LightGBM}} / \overline{\text{FR}}_{\tau,\text{QRF}} \qquad (10)$$

In general, $\text{FR}_{\tau,\text{skill}} \leq 1$, while for an excellent forecast at level $\tau$, we have $\overline{\text{FR}}_{\tau,\text{LightGBM}} = 0$ and $\text{FR}_{\tau,\text{skill}} = 1$. When $\text{FR}_{\tau,\text{skill}} > 0$, LightGBM outperforms quantile regression forests, while the higher the $\text{FR}_{\tau,\text{skill}}$, the better the LightGBM.

## 4. Results

Results of the applications of the algorithms are presented here, while those results will be discussed in detail in the next section, followed by respective explanations. Regarding the performance of the algorithm in the test set, we tested them at quantile levels $\tau \in \{0.5, 0.6, 0.7, 0.8, 0.9, 0.95, 0.97, 0.99, 0.999\}$. Recall that skill scores higher than 0 indicate that LightGBM outperforms QRF.

Regarding the case of the frequency skill score (recall the explanation of values of the frequency skill score $FR_{\tau,skill}$ after eq. (10)), presented in Figure 3a, the performance of both algorithms is almost equal for $\tau \leq 0.8$, while there is a fluctuation around 0 for $\tau \in (0.8, 0.95)$ and LightGBM outperforms QRF in higher quantile levels. Recall from Section 3.4, that the scoring functions related to frequencies are not consistent for a functional of interest, therefore a rigorous assessment of the algorithms is possible using the quantile scoring function. Skill score values for the quantile scoring function (recall the explanation of values of the quantile skill score $S_{\tau,skill}$ after eq. (8)) are presented in Figure 3b, where it seems that LightGBM outperforms QRF for $\tau \geq 0.97$, while the performances of both algorithms are approximately equal at lower quantile levels. In both cases of skill scores, the score increases with $\tau$ increasing when $\tau \geq 0.97$.

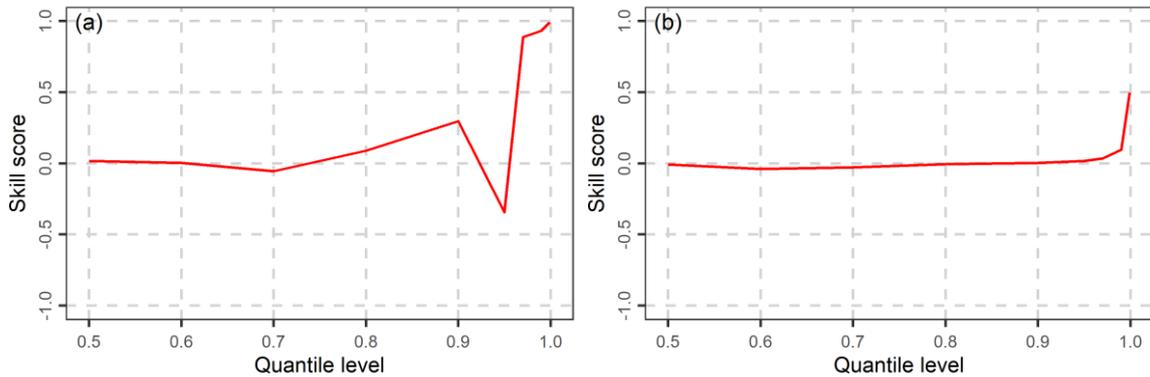

Figure 3. Skill scores for (a) frequencies and (b) quantile losses at different quantile levels $\tau$ for complete data in the test set.

It is of interest to understand how the algorithms behave when the observed values in the test set are equal to 0. Zero precipitation corresponds to approximately 72% of total daily observations, while intermittency in time and space is a dominant property of precipitation. Related skill scores for frequencies as well as quantile scoring functions are presented in Figure 4. Regarding frequencies, the performances of both algorithms are equal (Figure 4a). That is expected, since the algorithms issue always predictions (for the case of LightGBM, after adjustment of the predictions; see Section 3.2) that are equal or higher than 0; see also discussion in the next Section. However, QRF seems to outperform LightGBM for $\tau \geq 0.97$, regarding the quantile scoring function; see Figure 4b.

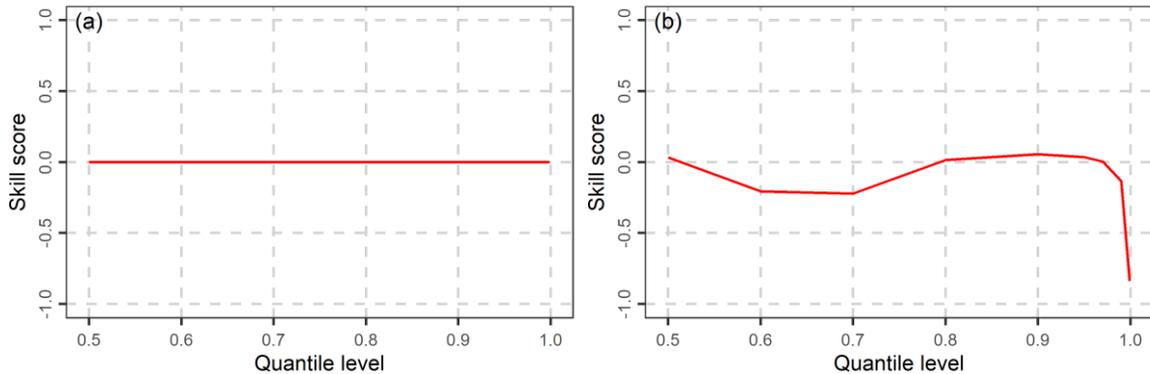

Figure 4. Skill scores for (a) frequencies and (b) quantile losses at different quantile levels for observed precipitation equal to zero in the test set.

The overall picture changes when testing the algorithms in observed precipitation higher than 0; see Figure 5. Here LightGBM seems to outperform QRF for $\tau \geq 0.95$ referring to both frequency and quantile scoring function based skill scores. In both cases of skill scores, the score increases with $\tau$ increasing when $\tau \geq 0.97$.

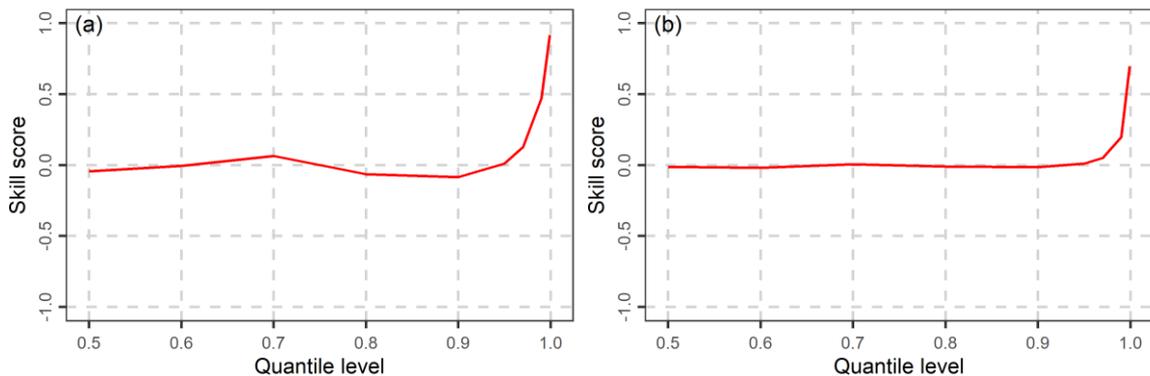

Figure 5. Skill scores for (a) frequencies and (b) quantile losses at different quantile levels for observed precipitation higher than zero in the test set.

It is also of interest to understand how the algorithms perform at each station separately; see Figure 6. Here, we examine the case of the quantile scoring function based skill score; recall that the quantile scoring function is consistent for the quantile functional. Stations with skill scores lower than –1 were removed from Figure 6. The reason is that, some skill score values were as low as –10 or less, which would create some artefacts in the representation of the results. The conclusions are not affected by the removal, given that skill scores are skewed, since they cannot exceed the value of 1, although they can be equal to −∞. Furthermore, we removed stations were both algorithms had a mean score equal to 0 (in which case the skill score is not defined). In Figure 6, we observe that the skill score increases as the quantile level $\tau \to 1$. Furthermore, we observe that the skill score varies between stations at the same quantile level, although the variation is relatively small. A notable departure of the skill scores from 0 is observed for quantile levels $\tau \geq 0.97$.

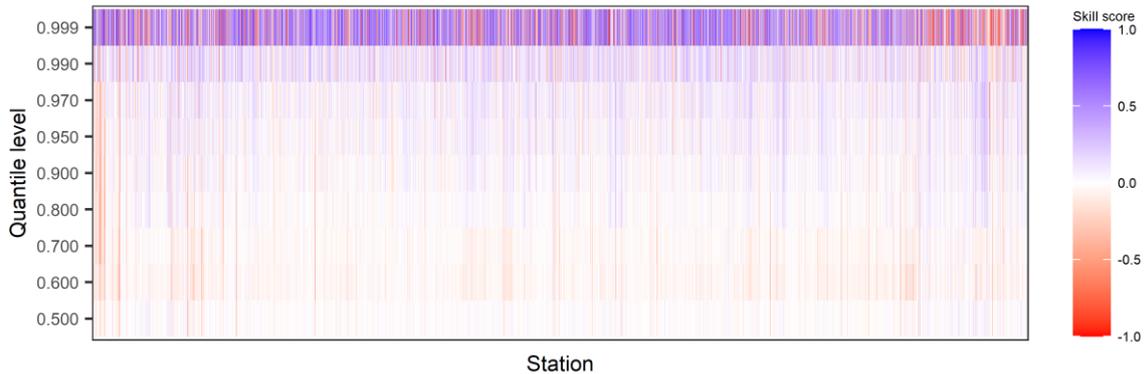

Figure 6. Heatmap of skill scores for quantile losses at different quantile levels and each station.

## 5. Discussion

Regarding the overall performance of the two methods, LightGBM is an algorithm particularly useful in large datasets with a high number of dimensions. Furthermore, given that it belongs to the highly parametrized family of boosting algorithms that are characterized by high flexibility, it is not surprising that it outperforms random forests on average. That is evident in the case of complete data in which LightGBM in general performs better when assessed with the quantile scoring function. However, LightGBM does not uniformly outperform QRF at all quantile levels. At lower quantile levels, the two algorithms seem to behave similarly, while at higher quantile levels LightGBM clearly outperforms QRF.

A possible explanation for the behaviour at lower quantile levels is based on the high proportion of zeros in the dataset. In particular, QRF is an algorithm based on bootstrapping therefore, it is possible to resample zero values. On the other hand, LightGBM is based on the minimization of the quantile scoring function that may be suboptimal when the dataset is highly intermittent. For instance, the median of the conditional distribution of precipitation should be zero (since the number of zero values in the dataset is higher than 50%). The quantile scoring function at level $\tau = 0.5$ is equivalent to the absolute error scoring function which in turn may not be suitable in cases that one should predict a median value of a probability distribution with mass at zero. Nevertheless, the overall performances of LightGBM and QRF at quantile level 0.5 remain similar. That may be due to that, while QRF can better predict zeros, they fail at a higher degree when they issue non-zero predictions.

At higher quantile levels, LightGBM clearly outperforms QRF with regards to all skill scores while the difference increases with increasing $\tau$, while the skill score tends to 1 as $\tau \to 1$. A possible explanation is that QRF cannot predict values that are not in the range of the training set (Hengl et al. 2018). The weakness becomes more pronounced at higher quantile levels, where high values in the training set become rarer and the algorithm tends to shift towards lower values. On the other hand, LightGBM is based on addition of base learners to previous errors and despite base learners being decision trees, it seems that it can better extrapolate beyond the range of the training set. Furthermore, the conditional probability distribution of precipitation at higher quantile levels seems to comply with regularity conditions, under which quantile scoring function's properties seem to hold.

While QRF outperforms LightGBM with regards to the quantile skill score at higher quantile levels when observed precipitation is zero, the inverse happens when observed precipitation is higher than zero. The performance of both algorithms in the complete test set favours LightGBM, since absolute values of quantile scores are lower in general when observed precipitation is zero compared to non-zero observed precipitation, consequently the largest part in the average score belongs to non-zero values.

Examining single algorithms is important to understand their properties at hand. However, combination of algorithms (ensemble learning in machine learning literature, Sagi and Rokach 2018), may result in higher improvements. That has been proved in practice using simple combinations (e.g. Papacharalampous and Tyralis 2020) or stacking (Tyralis et al. 2021) for predicting the mean functional of the conditional probability distribution. Combinations of algorithms for probabilistic prediction (e.g. stacking, Wolpert 1992, Lichtendahl et al. 2013, Yao et al. 2018), perhaps including spatial features (e.g. Papacharalampous and Tyralis 2022b) is a topic worth examining and has been proven successful in other hydrological applications (Papacharalampous et al. 2019, Tyralis et al. 2019a) as well as in merging gauged-based and satellite precipitation datasets (see Zandi et al. 2022, for the case of the mean functional). Predictions of extreme quantile based on extreme value theory is also worth examining, but it remains to assess in practice whether the conditional distribution in spatial settings is heavy-tailed. Nevertheless extremal quantile regression has been applied in post-processing applications in the time domain (Tyralis and Papacharalampous 2023). Assessing other algorithms, e.g. deep learning ones (LeCun et al. 2015, Schmidhuber 2015) in probabilistic predictions of precipitation might also be a topic worth examining.

## 6. Conclusions

We proposed issuing probabilistic predictions of daily precipitation in spatial settings of merging gauge-based measurements and satellite precipitation products using LightGBM. LightGBM outperforms the state-of-the-art in such settings quantile regression forests (a variant of random forests) when predicting extreme quantiles of the conditional probability distribution of the response variable, while both algorithms show similar performance when predicting quantiles at the centre of the probability distribution. The difference in the performance of the methods increases in favour of LightGBM as the quantile level (at which the methods are compared) increases and tends to 1. Confidence on the results is built through the comparison of the algorithms in a large dataset that includes observed precipitation in the contiguous US.

An intuitive explanation of the results is also provided, according to which LightGBM can better predict extreme quantiles due to the inability of random forests to extrapolate beyond the range of the training set combined with the improved ability of LightGBM to issue accurate predictions due to its structure. On the other hand, quantile regression forests are equal to LightGBM when predicting quantiles at the centre of the conditional probability distribution, due to the highly intermittent nature of precipitation, combined with their bootstrap-based structure, that seems to be more suitable in this case compared to algorithm structures that are based on the quantile scoring function.

**Conflicts of interest:** The authors declare no conflict of interest.

**Author contributions:** HT and GP conceptualized and designed the work with input from ND and AD. HT and GP performed the analyses and visualizations, and wrote the first draft, which was commented and enriched with new text, interpretations and discussions by ND and AD.

**Funding:** This work was conducted in the context of the research project BETTER RAIN (BEnefiTTing from machine lEarning algoRithms and concepts for correcting satellite RAINfall products). This research project was supported by the Hellenic Foundation for Research and Innovation (H.F.R.I.) under the "3rd Call for H.F.R.I. Research Projects to support Post-Doctoral Researchers" (Project Number: 7368).

**Appendix A     Statistical software**

We used the `R` programming language (R Core Team 2022) to implement the algorithms, and to report and visualize the results.

For data processing and visualizations, we used the contributed `R` packages `data.table` (Dowle and Srinivasan 2022), `elevatr` (Hollister 2022), `ncdf4` (Pierce 2021), `rgdal` (Bivand et al. 2022), `sf` (Pebesma 2018, 2022), `spdep` (Bivand 2022, Bivand and Wong 2018, Bivand et al. 2013), `tidyverse` (Wickham et al. 2019, Wickham 2022).

The algorithms were implemented by using the contributed `R` packages `ranger` (Wright 2022, Wright and Ziegler 2017), `lightgbm` (Shi et al. 2022).

The performance metrics were computed by implementing the contributed `R` package `scoringfunctions` (Tyralis and Papacharalampous 2022a, 2022b).

Reports were produced by using the contributed `R` packages `devtools` (Wickham et al. 2022), `knitr` (Xie 2014, 2015, 2022), `rmarkdown` (Allaire et al. 2022, Xie et al. 2018, 2020).